# Comment on "Proximity breakdown of hydrides in superconducting niobium cavities"

G Ciovati, P Dhakal and G R Myneni
Thomas Jefferson National Accelerator Facility, Newport News, VA 23606, USA

According to the authors of this article, a reduction of vacancies concentration at the Nb surface occurs after 120 °C bake, resulting from the dissociation of vacancies-hydrogen complexes, is proposed as an explanation for the reduction of the high-field Q-slope after baking in SRF cavities made of bulk Nb. Nevertheless, recent positron annihilation spectroscopy (PAS) data measured on Nb samples which underwent similar treatments to those of SRF cavities do not support this hypothesis, as they showed an increase of vacancies concentration after baking at 145 °C [1].

Furthermore, the authors imply that a very high concentration of vacancies near the Nb surface is still present after vacuum heat-treatments (600-800 °C). On the other hand, PAS data clearly show that the vacancy concentration in deformed or irradiated Nb samples is greatly reduced after vacuum heat treatments above 700 °C [2], approaching that of a "defect-free" Nb after vacuum heat treatment at 1000 °C [3].

Regarding the curve fitting of $R_s(H_{peak})$ data, the parameter $R_0$ represents the portion of the surface resistance unaffected by the presence of hydrides, according to the model. As such, one would expect similar values of $R_0$ for both data sets in Figs. 2 and 3, yet, the value of $R_0$ for the data in Fig. 3 is about a factor of three higher than that in Fig. 2. The parameter σ in Eq. (3) corresponds to the width of the distribution of the inverse of the breakdown field, not the width of the breakdown field distribution, as stated in the article.

## References

[1]  Visentin M, Barthe M F, Moineau V and Desgardin P 2010 *Phys. Rev. ST Accel. Beams* **13** 052002.
[2]  Naidu S V, Gupta A Sen and Sen P 1987 *J. Nucl. Mater.* **148** 86-91.
[3]  Cizek J, Prochazka I, Brauer G, Anwand W, Gemma R, Nikitin E, Kirchheim R and Pundt A 2008 *Acta Phys. Pol. A* **113** 1293-99.